\documentclass[10pt,conference]{IEEEtran}
\usepackage{courier}
\usepackage{url}
\usepackage{balance}
\usepackage{enumitem}
\usepackage{caption}	
\usepackage{multirow}
\usepackage{multicol}
\usepackage{amsmath}
\usepackage{float}
\usepackage[normalem]{ulem}
\usepackage{xspace}
\usepackage{xcolor}
\usepackage{listings}
\usepackage{subcaption}
\usepackage{hyperref}
\usepackage{graphicx}
\usepackage{tabularx}

\usepackage{hyperref}
\include{macro}

\newcommand{\Guigle}{{\sc Guigle}\xspace}
\newcommand{\Guigles}{{\sc Guigle's~}\xspace}
\newcommand{\ReDraw}{{\sc ReDraw}\xspace}
\newcommand{\CrashScope}{{\sc CrashScope}\xspace}
\newcommand{\MonkeyLab}{{\sc MonkeyLab}\xspace}
\newcommand{\ie}{\textit{i.e.,}\xspace}
\newcommand{\eg}{\textit{e.g.,}\xspace}

\newcommand{\etal}{\textit{et al.}\xspace}

\newcommand{\figref}[1]{Fig.~\ref{#1}\xspace}

\newcommand{\totalAfterFilter}{7,654\xspace} 
\newcommand{\totalIndexed}{5,416\xspace} 
\newcommand{\totalImagesIndexed}{12,051\xspace}

\begin{document}
%
\title{Guigle: A GUI Search Engine for Android Apps}


\author{\IEEEauthorblockN{Carlos Bernal-C\'ardenas, Kevin Moran, Michele Tufano,\\ Zichang Liu, Linyong Nan, Zhehan Shi, and Denys Poshyvanyk}
\IEEEauthorblockA{Department of Computer Science\\
College of William \& Mary\\
Williamsburg, VA\\
Email: \{cebernal, kpmoran, mtufano, lzcemma, lnan, zshi01, denys\}@cs.wm.edu}}

\maketitle

\begin{abstract}
The process of developing a mobile application typically starts with the ideation and conceptualization of its user interface. This concept is then translated into a set of mock-ups to help determine how well the user interface embodies the intended features of the app. After the creation of mock-ups developers then translate it into an app that runs in a mobile device. In this paper we propose an approach, called \Guigle, that aims to facilitate the process of \textit{conceptualizing} the user interface of an app through GUI search.  \Guigle indexes GUI images and metadata extracted using automated dynamic analysis on a large corpora of apps extracted from Google Play. To perform a search, our approach uses information from text displayed on a screen, user interface components, the app name, and screen color palettes to retrieve relevant screens given a query. Furthermore, we provide a lightweight query language that allows for intuitive search of screens. We evaluate \Guigle with real users and found that, on average, $68.8\%$ of returned screens were relevant to the specified query. Additionally, users found the various different features of \Guigle useful, indicating that our search engine provides an intuitive user experience. Finally, users agree that the information presented by \Guigle is useful in conceptualizing the design of new screens for applications.

\noindent \textbf{Video URL: \url{https://youtu.be/hqUuuMMj2BU}}

\end{abstract}

\vspace{-0.2cm}
\section{Introduction \& Motivation}
\label{sec:introduction}

Mobile devices and apps have an important impact on the everyday lives of people around the world. This impact stems from the ability of these apps to enable a range of tasks, from simple chores such as calculating a tip for a meal to more complex activities. These tasks are enabled by the rich ecosystem of ``apps'' available on mobile devices. However, before developers publish their apps to a marketplace such as Apple's App Store~\cite{AppleStore} or Google Play~\cite{GP}, they must endeavor to build an app following best practices for mobile software development. This process starts with the ideation and conceptualization of the requirements and user interface of the app. The process then proceeds to the creation of a set of screen mock-ups that delineate the graphical user interface (GUI).  User interface and user experience (UI/UX) designers typically iterate these mock-ups until all the features are captured in the GUI. Once the final design is ready, programmers translate the mock-up \NEW{(typically created in software like Sketch~\cite{sketch})} and resources provided by designers into a runnable app. After validating that the app successfully passes a suite of tests, it is published on a market. 

	One of the most difficult parts of this process is designing an intuitive GUI and creating an effective mock-up to capture all required functionality of an app. \NEW{In this paper we focus on improving this design task by facilitating the process of finding example app screens that are relevant to a query formulated according to app design requirements}. To accomplish this we have designed and implemented \Guigle, a search engine that assists users in finding relevant screenshots of apps to help aid in GUI-design. \Guigle indexes a large corpus of $5k$ apps consisting of over $12k$ screens and enables advanced searches using Natural Language (NL) queries and result filtering according to metadata such as color palettes, screen types (\eg settings screen), and GUI-component types (\eg returning screens that include progress bars or buttons).

\begin{figure*}[t]
    \begin{center}
	\vspace{-0.75cm}
		\includegraphics[width=\linewidth]{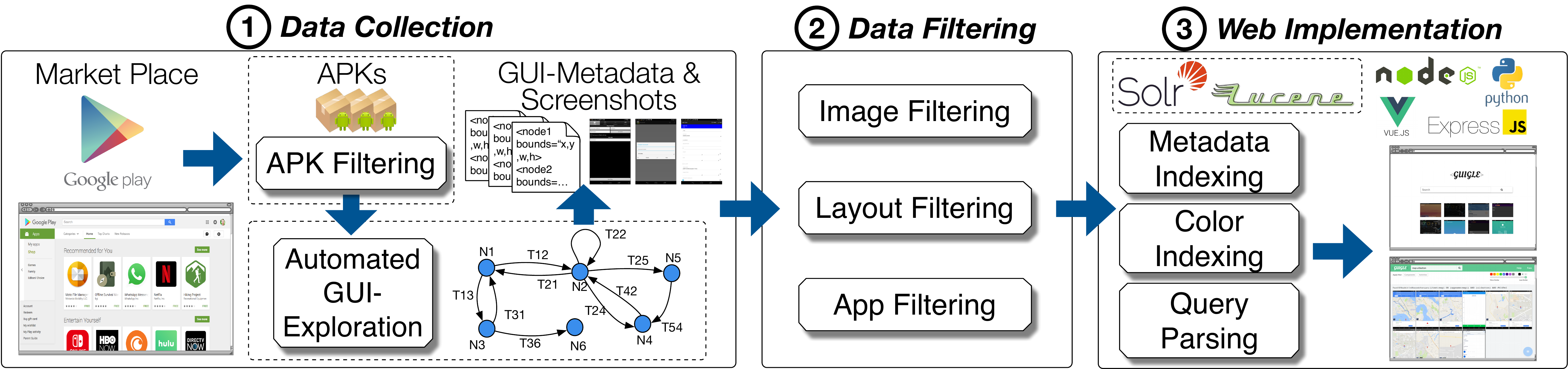}
		\vspace{-0.45cm}
        \caption{The \Guigle approach and components}
        \label{fig:approach}
        \vspace{-0.75cm}
    \end{center}
\end{figure*}

\Guigle represents a significant departure from and improvement over existing image search engines that help to illustrate its novelty. Typically, search engines such as Google Image Search~\cite{google-images} utilize computer vision techniques and indexed metadata from the web in order to return relevant images in relation to a user's NL query. However, this is a more general search tool that does not allow for detailed searches of a large index of Android application screens and lacks capabilities for filtering searches by screen or GUI-component type. The most closely related approach to our tool is a recently published framework called GUIFetch~\cite{Behrang:MOBILESoft18} that is capable of retrieving example code snippets from relevant open-source applications using a design mock-up and keywords as a query. While the GUIFetch approach represents a promising technique for helping developers to translate an existing mock-up into code by retrieving implementation examples, it does little to support designers and developers during the early stages of app GUI conceptualization as it requires a mock-up as input. Conversely, \Guigle is a complementary approach implementing a lightweight way to quickly search a large number of app GUIs and visually inspect the results, facilitating the process of conceptualizing a GUI based on the collective design patterns of retrieved screens. Furthermore, \Guigle supports a set of simple but powerful search query formulations that allow users to quickly discover app screens relevant to highly specific concepts (\ie via screen and GUI-component type filtering), that GUIFetch does not support. Finally, \Guigle is able to index a large number Google Play apps without the need for access to source code, whereas GUIFetch needs access to source code and thus is limited to a smaller set of open source apps.

\Guigles contributions can be summarized as follows:
\vspace{-0.1cm}
\begin{itemize}
	\item{\NEW{A technique for indexing a large corpus of mobile app screens combining image and metadata processing;}}
	\item{A lightweight query language allowing for intuitive search of relevant app screens;}
	\item{\NEW{A set of user controls for filtering and refining queries based upon component type, screen type, and color;}}
	\item{\NEW{A publicly available implementation of \Guigle, embodying the above techniques~\cite{guigle-online}.}} \end{itemize}
\vspace{-0.3cm}

\section{Approach}
\label{sec:approach}

\figref{fig:approach} provides an overview of \Guigle and its three main components. The first component is responsible for downloading Android APKs from Google Play, automatically executing them via systematic exploration, and collecting screenshots and the associated GUI-hierarchy metadata. Subsequently, a data filtering step discards uninteresting GUIs (i.e., blank screens or those with very few components). Finally, the GUI metadata and color information is indexed using Lucene~\cite{lucene}. In this section, we describe Guigle's components in detail.

\vspace{-0.2cm}
\subsection{Data Collection}
\NEW{The first major component of \Guigle, collects Android APKs and extracts the data that enables the creation of a GUI screenshot search engine. To enable GUI-search across a large number of screens, we utilized the dataset derived for the \ReDraw~\cite{Moran:TSE18} approach for automated app prototyping, with some additional post-processing to ensure a high-quality index of screens. This dataset consists of \totalAfterFilter apps after the removal of non-native apps. For the remainder of this section \& Sec. \ref{subsec:data-filtering}, we briefly review the methodology used to collect the ReDraw dataset, and detail the processing steps utilized to derive \Guigle. We refer the reader to \cite{Moran:TSE18} for further details.}

\subsubsection{Execution Engine}
In order to collect screenshots and metadata of multiple activities from the downloaded apps, each app was systematically explored in a Depth-First-Search (DFS) fashion, using the systematic input generation approach developed as part of our prior work on \CrashScope and \MonkeyLab~\cite{Linares:MSR15, Moran:ICST16}. During this exploration, each GUI event generated (\eg click of a button) produced a screenshot and \texttt{\small xml dump} (via the \texttt{\small uiautomator tool}) that contains information regarding the hierarchy of GUI-components on the screen. Once the exploration of an app was completed, only a set of unique screens were selected among the top-6 most frequently ``visited'' screens of the app. The rationale behind this selection strategy is that frequent screens (\ie appearing multiple times during the systematic exploration) correspond to the most frequent activities used in an app, and thus characterize the app's functionality. Further details regarding the automated app exploration can be found in the paper describing the full ReDraw approach \cite{Moran:TSE18}. 

\subsection{Data Filtering}
\label{subsec:data-filtering}
To ensure the quality of our extracted dataset we systematically removed low-quality screens and sampled a statistically significant subset of screens for manual validation.

\subsubsection{Image Filtering}
One quality issue we identified in our initial dataset was related to collected screenshots of the Android home screen caused by apps failing to properly launch or restarting during the automated execution. We discarded the screenshots whose \texttt{\small xml dump} file contained \texttt{\small com.an\-droid.launcher} in the string representing the package name, indicating a home screen. We also observed screens that included an \textit{\small overlay}, often meant to provide an overview of the app functionality or to indicate how to exit from an app's ``full screen'' mode. To identify and remove these screens, we applied color histogram analysis for distinguishing repetitious color patterns focused upon screen borders, since these areas were the most common ones to find the overlays.

\subsubsection{Layout Filtering}
Additionally, our initial dataset covered screens that only included GUI \textit{containers} or GUI-components meant to group other GUI-components only \eg \texttt{\small View, GridLayout} among other containers. We parsed all the \texttt{\small xml dump} files and discarded all the screens that contain only these types of GUI-components, since they do not provide relevant information in posterior steps.

\subsubsection{Google Play Description Filtering}
\Guigle includes information provided from Google Play to provide a better user experience. However, after downloading and executing app from the Google Play and and attempting to link app descriptions at a later date, we discovered that some apps were removed or inaccessible. This was likely due to apps not satisfying Goole Play terms and conditions or being removed by developers. Therefore, we discarded these applications from our dataset as we considered this apps to not be relevant for \Guigle resulting on a total of \totalIndexed unique apps.

\subsection{Web Implementation}

We implemented \Guigle as a web application in Python using different tools to provide performant query responses.

\subsubsection{Metadata Indexing}
The first step involved the indexing of data included in the \texttt{\small xml dump} files. We used Apache Lucene~\cite{lucene} to index attributes such as the \texttt{\small app name}, \texttt{\small component type}, and \texttt{\small text} of each GUI-component mapping them to a corresponding field of Lucene's document. This enabled multi-field search and allows for the creation of complex queries to assist the user in obtaining relevant results.

\subsubsection{Color Indexing}
In order to perform search by color, we extracted the top-6 colors from each of the screenshots using the \textit{colorgram}~\cite{colorgram} Python library. This library allowed us to extract a simplified 6 color palette from the original image and by extracting color groups, and averaging the color values for similar groups. For each color extracted, we transformed the RGB value into the \textit{Hue}, \textit{Saturation}, and \textit{Lightness} (HSL) color space. Moreover, we extracted the proportion in terms of the percentage on each of the top-6 colors for each screenshot.

\subsubsection{Query Parsing}
We implemented a query parser to provide a more user-friendly search experience. This parser follows a four step process. The first step, applies preprocessing to clean up the query, for instance by removing additional spaces between words. The second step detects and classifies tokens in the query into one of four categories using keywords. If the tokens correspond to any value in the suggestions' list then the corresponding category is used in the final query. The keywords are (i) \textit{color}, which uses a standard list of colors widely used in web browsers  as suggestions, or additionally, the user can use any hex value to be more specific; (ii) \textit{ui}, which uses the type of GUI-components as suggestions (\ie components' class names); (iii) \textit{appname} and (iv) \textit{text}, which are used for strings not classified as either \textit{color} or \textit{ui} and are used to search the application names and text displayed on components respectively.

	The third step handles logical operators between tokens. Therefore, if the user does not specify either \textit{AND} or \textit{OR}, by default the \textit{AND} logical operator is applied between the pair \textit{$<$keyword:value$>$}. Conversely, if users specify the logical operator then the parser keeps it. To avoid ambiguous queries the user has to add parentheses in the cases in which the operators \textit{AND} and \textit{OR} are used in the same query. To provide an example of how our query parser operates, consider the following query \texttt{\small red edittext pizza}, for which the parser would output \texttt{\small color:red AND ui:edittext AND (text:pizza OR appname: pizza)}.~In this case \texttt{red} is categorized as \textit{color}, \texttt{\small edittext} is identified as \textit{ui}, and since \texttt{\small pizza} is not identified as either \textit{color} nor \textit{ui} it assigns \textit{appname} and \textit{text} keywords with the \textit{OR} connector on the final query. 

	The last step uses a predefined query filters for types of GUI-components and activities to speed up the search on simple queries. Additionally, users can use a color picker to filter screens for any query according to manually specified screen colors. Moreover, we provide a slide bar that can be used to specify the maximum difference that can be considered for a color to be close to another one. This range is used for each of the color components of the HSL color space. 

\subsubsection{Guigle User Experience (UX)} \Guigle provides dynamic search suggestions when the user enters keywords in order to speed up the process of formulating queries. Once results are returned, users can use \Guigle to get detailed information for each screenshot by clicking on it. This detailed result view provides information such as name of the app, top-6 colors sorted by proportion, list of GUI-components, link to Google Play, similar screens based on the GUI-components, and all other screens of the same app. Additionally, users can favorite screenshots and access them later for quick inspection.

To facilitate the implementation of our search engine, we relied upon Apache Solr~\cite{solr} that helps to expose Lucene's functionality with \texttt{\small RESTful} web services. This allows for seamless integration with web frameworks such as Node Js, Express Js, and Vue Js. The combination of these frameworks enables capabilities that facilitate Guigle's user experience and allows for a responsive web application that can scale to different screen sizes or browser viewport widths.

\section{Evaluation}
\label{sec:evaluation}

\subsection{Study Design}
The \textit{goal} of our study is to evaluate the usefulness of \Guigle in terms of (i) its \textit{effectiveness} in retrieving relevant screenshots and \NEW{(ii) the \textit{usability} of exercising search features.}

	Concerning the study \textit{methodology}, we created a survey structured in four sections: (i) demographic questions; (ii) a set of tasks which simulate the scenarios where a developer/designer is seeking inspiration from other app's screenshots; (iii) quantitative questions about the usability of \Guigle; and (iv) qualitative questions related to the UX of \Guigle.

	The first section of the survey aims to help understand the demographic makeup of our participants, whereas the second section aims to evaluate whether each of the screens, yielded by a \Guigle search, is relevant to the task the user is supposed to complete. To quantify this, we based our survey on common procedures used to evaluate search engines ~\cite{McMillan:TSE12}. The main measure we used to evaluate the \textit{effectiveness} of \Guigle is the precision defined as $P$=$relevant/retrieved$, where $relevant$ constitutes the screenshots considered pertinent in the search while $retrieved$ represents the number of total results returned by the search. Furthermore, we asked the users to evaluate the relevancy of the top-10 screenshots of each query created by the participants based on the task, leading to our $retrieved$ variable being fixed at 10. The third survey section evaluates usability based on six questions where the user can express her usability assessment on a 5 point likert-scale.  This allowed us to evaluate \Guigle's effectiveness in the context of each task. We derived the questions of our study based on SUS usability scale by Brooke \cite{Brooke:96}. Finally, the fourth section collects qualitative feedback using four open questions that helped us to gather additional information on \Guigles UX. These questions were derived according to the user experience honeycomb by Morville  \cite{Morville:04}.

\NEW{The \textit{context} of our study comprises 13 developers who completed our survey.} Moreover, we indexed a total of \totalImagesIndexed documents with Lucene which includes $\mathtt{\sim}$12k screenshots from a total of \totalIndexed unique apps.

\subsection{Results}
\vspace{-0.1cm}
We surveyed 13 developers with an average of about 4 years of experience in software development and 10 months in mobile development. Our participants came from a variety of backgrounds, as indicated by their highest obtained degree including:  23\% with high school, 62\% with bachelors, 7.5\% with masters, and 7.5\% with PhD degree completed.

\begin{figure}
	\vspace{-0.4cm}
	\centering
	\begin{subfigure}{.5\linewidth}
		\includegraphics[width=\linewidth]{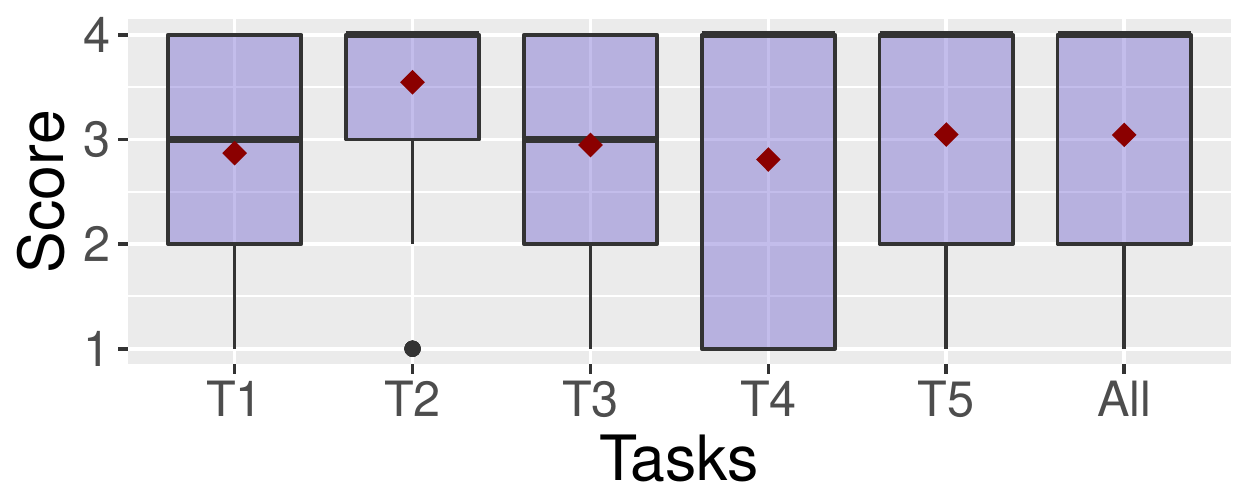}
		\vspace{-0.5cm}
		\caption{Score confidence}
		\label{fig:score}
	\end{subfigure}%
	\begin{subfigure}{.5\linewidth}
		\includegraphics[width=\linewidth]{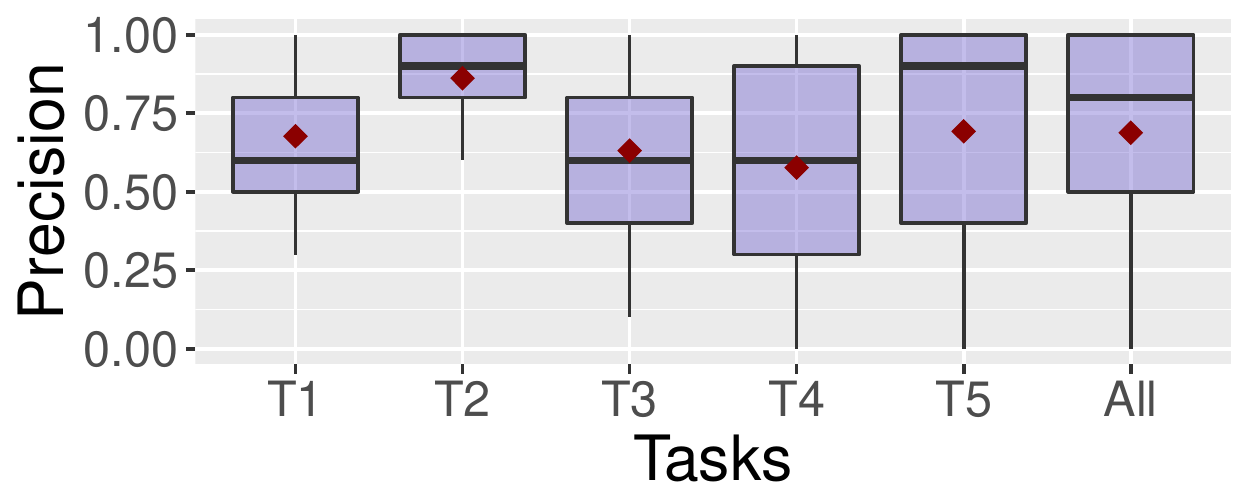}
		\vspace{-0.5cm}
		\caption{Precision}
		\label{fig:precision}
	\end{subfigure}
	\vspace{-0.1cm}
	\caption{Score confidence and Precision per task across participants of the survey}
	\label{fig:fig}
\end{figure}

\subsubsection{Precision Analysis}
In terms of the \textit{effectivness}, the average precision regarding the relevance of returned screens is 68.8\%, as shown in \figref{fig:precision} where the \textit{x} axis presents all the tasks and the \textit{y} axis shows the percentage in terms of precision. The results suggest that \Guigle is able to find relevant screenshots for a given query with an average confidence score of 3 (\ie mostly relevant). Additionally, this may indicate that the attributes such as \textit{app name}, \textit{GUI component type}, \textit{text}, and \textit{color} are useful for searching screenshots.

\subsubsection{Quantitative Analysis}
\begin{figure}
	\begin{center}
		\vspace{-0.5cm}
		\includegraphics[width=\linewidth]{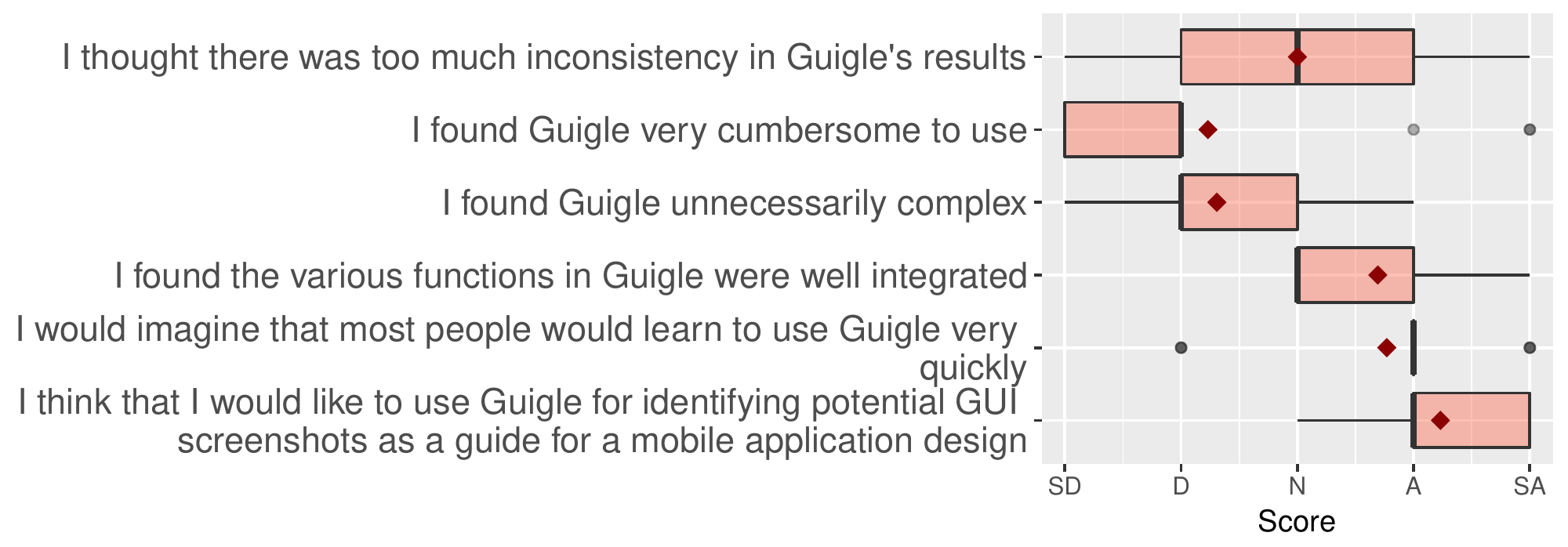}
		\vspace{-0.6cm}
		\caption{Survey section 3 - quantitative analysis}
		\label{fig:quantitative}
		\vspace{-0.8cm}
	\end{center}
\end{figure}

The results of the survey for the usability section are presented in \figref{fig:quantitative} where the \textit{x} axis reports the score for the likert-scale whereas \textit{y} axis shows the set of questions. It is worth noting that the survey included three negative questions (\ie the lower the better) and 3 positive questions (\ie the higher the better) to avoid bias in the responses. As a result, we found that users agree with positive questions suggesting that \Guigle exhibits intuitive usability. On the other hand, people disagreed with the negative questions reinforcing this observation. However, study participants reported a neutral sentiment concerning the consistency of \Guigles returned results, indicating the possibility of occasional irrelevant screens in a result set.

\subsubsection{Qualitative Analysis}
\begin{figure}
	\begin{center}
	\vspace{-0.4cm}
		\includegraphics[width=.9\linewidth]{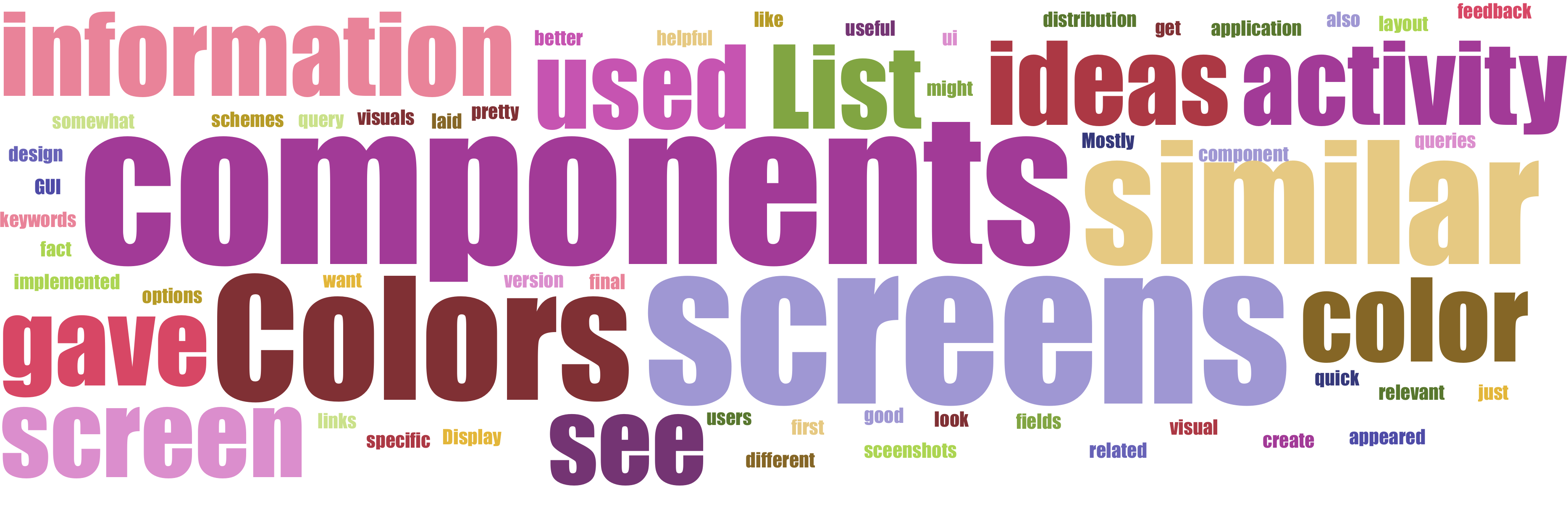}
		\vspace{-0.1cm}
		\caption{Survey section 4 -  qualitative analysis}
		\vspace{-0.7cm}
		\label{fig:qualitative}
	\end{center}
\end{figure}

The qualitative section of the survey focused on open questions related to the usability. Due to space limitations, in \figref{fig:qualitative} we present a cloud word from answers to the question: \textit{What information did you find useful from these screenshots?}. This suggests that similar screens from the same app are relevant for the user. \Guigle's capability of search-by-color also appears to be considered very useful by the participants, which allow them to search apps with a given color theme/palette. However, some participants mentioned that some of the searches did not rank higher screens having larger areas of the selected color. This might be due to the fact that, currently, \Guigle does not rely on the proportion of the color to rank the resulting screens.

\section{Demo Remarks \& Future Work}
\label{sec:conclusion}

In this demo, we presented \Guigle, a GUI search engine for Android app screenshots which supports queries for searching according to (i) the \textit{app name}, (ii) GUI-component \textit{text}, (iii) GUI-component \textit{type}, and (iv) screen \textit{color}. Users can inspect each of the retrieved screens and obtain detailed information such as the list of GUI-components, the name and link to the Google Play store of the belonging app, additional screens of the same app, top-6 colors sorted by the proportion of the color in the screen, and other similar screens. \Guigle was evaluated by 13 developers on an online survey. 
The results suggest that \Guigle is effective in retrieving relevant screens while providing an intuitive user experience through its web interface.

In the future, we plan to add support for more complex queries to enable more \NEW{robust} searches. This includes the possibility of creating queries that consider the hierarchy of components and consider proportion of colors to better rank the screenshots. Furthermore, \Guigle can be envisioned as a starting point for providing developers with app skeletons for closed source apps. This could be done by leveraging approaches that generate GUI code based on screenshots such as REMAUI~\cite{Nguyen:ASE15}, \ReDraw~\cite{Moran:TSE18}, and Chen \etal~\cite{Chen:ICSE18}.

%
\IEEEpeerreviewmaketitle


\balance
\bibliographystyle{abbrv}
\bibliography{ms.bib}

\end{document}